\documentstyle[aps,epsf,twocolumn]{revtex}

\newcommand{\be}{\begin{eqnarray}}
\newcommand{\ee}{\end{eqnarray}}

\begin{document}

\title{ Instanton size distribution: repulsion or
 the infrared fixed point?}

\author{E.V.~Shuryak \\Department of Physics,\\ State University of New York
 at Stony Brook, Stony Brook,\\ New York 11794, USA}
\maketitle

\begin{abstract}
  We discuss  available information about
the instanton size distribution $d(\rho)$, which comes from lattice simulations
and the interacting instanton liquid model. Not only they are remarkably
consistent, but one can also reproduce $d(\rho)$ with an alternative idea,
based on the
 infrared fixed point. We show, that
lattice non-perturbative beta-function (for $bare$ charge renormalization)
also suggests such a hint. We also discuss
whether  it is possible to reconcile the
instanton physics with reasonable large-number-of-colors limit, concluding
that our fits
for  SU(2) and
SU(3) theories are actually very close to
the critical value $8\pi^2/g^2= N_c*5.4$ when it is the case.
\end{abstract}
\narrowtext

   Instantons were discovered 20 years ago \cite{BPST}, but only
now their significant role in strong interaction physics
is being widely recognized.  A large set ($\sim$ 40)
 of point-to-point correlation
functions was calculated  in the  simplest ``instanton liquid"
model \cite{Shuryak_1982}, both numerically
\cite{RILM} and (with certain approximations) analytically
\cite{DP_cor}. The results are truly spectacular:
not only masses and coupling constants of lowest hadronic states
(including $\pi,\sigma,\rho,A_1,N,\Delta$ and others) are reproduced
without additional parameters,
 but the whole correlation functions happen to be  in agreement
with those extracted from phenomenology \cite{Shuryak_cor} and from
lattice simulations
\cite{Negele}. This implies that instanton-induced
't Hooft interaction \cite{tHooft} between light quarks is dominating
the interquark forces.
Recently, glueballs were added to the list \cite{SS_glue}, with the conclusion
that their specific features in the instanton model
(such as small mass and especially small $size$ of
the scalar glueball) can $quantitatively$ explain available lattice data.

 This agreement is not accidental, as can be seen from
the fact that
the ``instanton liquid" itself was
``distilled" from lattice configurations by the ``cooling" method
\cite{PV,CGHN_94,MS}.
In spite of the (nearly complete) loss of confinement and of the perturbative
gluons, this operation  does
not affect much  these correlators (and hadrons)
 \cite{CGHN_94}. Furthermore, the
main parameters (the average instantons size $\bar\rho$  and
instanton spacing R) were found to be
inside 10\% the same as those suggested in 1982
\cite{Shuryak_1982}, namely $\bar\rho\approx 1/3 fm,R\approx 1 fm$.

The topic of this work is  the $shape$
of the instanton (plus antiinstanton)size distribution
$d(\rho)$,  related with a number of important issues,
both practical and theoretical ones.
  With all those   advances at the phenomenological front, we are still
lacking  answers to many major questions. One of them, to be
discussed in this letter, is:
Why  are  $large-size$ instantons  absent in the QCD vacuum?
  Alternative explanations are:
(i) they are suppressed by $repulsive$ interaction between instantons;
(ii) the {\it  higher-order effects} lead to charge renormalization
so that their actions are large;
(iii) $confinement$ effects screen their gluoelectric fields.
In this Letter we will go through this list, and will show that the first two
are still strongly competitive.

   The first idea is the
basis of the
'interacting instanton liquid model' (IILM), formulated as a particular
statistical model amenable for numerical simulations
\cite{Shuryak_improve,SV}, its
partition function (for $N/2$ instantons
and $N/2$ anti-instantons) is generally given by
\be
Z = \int \prod_{I=1}^N d\Omega_I d_0(\rho_I)
\exp[-\sum_{I < J} S^{int}_{IJ}]
\ee
where $\Omega_I$, denote the orientation, position and
the size of pseudoparticle $I$,
$d_0(\rho)$ here corresponds to $non-interacting$ instantons, with the
  gluonic interactions included separately.
(The quark-related
ones are not
discussed in this letter.)

The simplest ``hard core" model was introduced in
\cite{IM}. Then it was shown that for the simplest ``sum ansatz"  \cite{DP}
such a repulsion actually exists,
even slightly stronger than it is needed.
Some defects of this ansatz was cured
in \cite{Shuryak_improve}, where the
improved trial function known as ``ratio ansatz" was proposed: this
reproduces phenomenological parameters of the ``instanton liquid" well.
However, numerical solution \cite{V_streamline}
 of the ``streamline" equation \cite{BY} have shown that the
instanton-antiinstanton valley leads
continuously to zero fields.
  Thus, it seems that the original hopes to stabilize the ensemble
at the purely classical level \cite{DP} are not
fulfilled. Presumably quantum effects (especially
subtraction of  perturbative contributions, relevant for close
instanton-anti-instanton pairs with a strongly attractive interaction) will
generate the effective repulsion (see also \cite{DP_quantum}).
Meanwhile, in IILM-based recent
studies  \cite{SS_glue,SV_screening} the streamline
interaction is supplemented by a repulsive core, with the radius fitted to
the value of the gluon and quark condensates.

  In Fig.1 (open points)
we show  the resulting instanton
size distribution in IILM,
for the pure gauge SU(2) (a) and SU(3) (b)
theories. In the former case they are compared to first available lattice data
\cite{MS} (closed points). They are
 very similar,
except at small sizes  (instantons with the radius $\rho\sim 1$
 in lattice units ``fall through the lattice" during ``cooling").
At $large$ sizes both
lattice data and ``interacting liquid" give
virtually identical results.

\begin{figure}
\begin{center}
\leavevmode
\epsffile{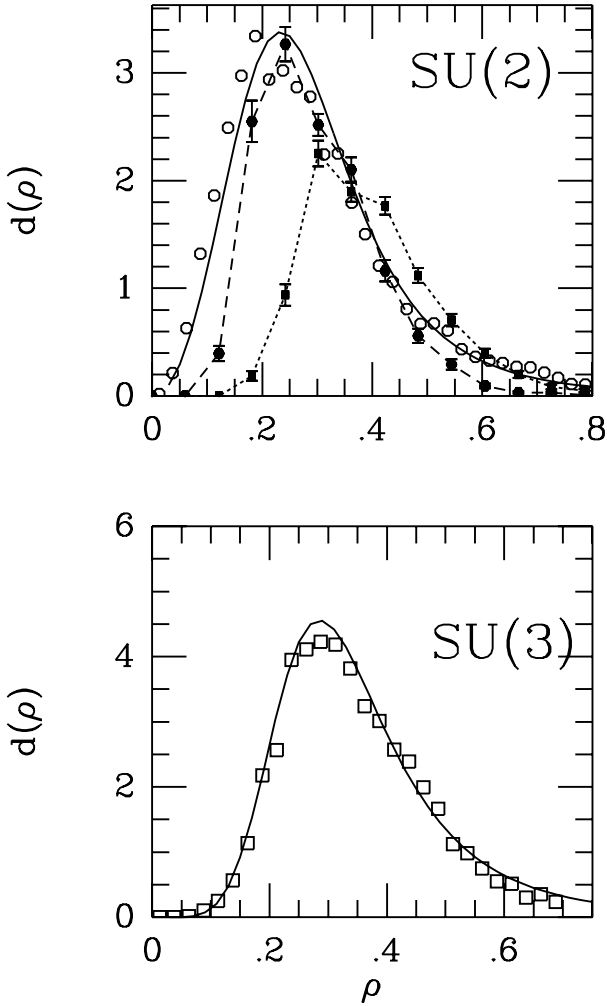}
\end{center}
\caption{The
 instanton size distribution in d=4 SU(2) (a) and  SU(3) (b) gauge theories.
 The open points correspond to interacting
instanton liquid", and the
closed ones are
   lattice results By Michael and Spenser, $16^4,4/g^2=2.4$ for squares and
$24^4,4/g^2=2.5$ for dots (the dotted and dashed lines just guide the eye).
 Solid curves correspond to the parametrization discussed in the text.
Units are in ``femtometers", defined for lattice data by
the scalar glueball mass defined as
$m_{0^+}=1.7 ``GeV"$; and by $1/\Lambda_{PV}$ for the instanton model.
}
\end{figure}

   It is well known that (provided the action
is large enough
$S_{eff}>>1$)
one can use the semiclassical
theory and derive the following
 generic expression
\be
d(\rho)= {2 C_I \over \rho^{(d+1)}} S^{[N_{zm}/2]}(\rho) exp[-S(\rho)]
\ee
where 2  accounts for instantons and anti-instantons, d is
the space-time dimension, and $N_{zm}$ is the number of zero modes ($4N_c$ for
Yang-Mills fields).
 The value of the constant (which depends
on renormalization method) was determined for the Yang-Mills instantons in
the classical work by 't Hooft
\cite{tHooft}. In order to make our discussion below
 simpler, we
use $C_I=0.466/ [(N_c-1) !(N_c-2) !]$, absorbing the $N_c$ dependent
 factor
into  a new Lambda parameter
$\Lambda_{inst}=
0.632 \Lambda_{Pauli-Villars} = 0.657 \Lambda_{\bar M \bar S}$.)

Multiple arguments in favor
of existence of the {\it infrared fixed point}
(or ``freezing" of the coupling constant) have been many times made
before, based on a variety of
phenomenological models
 (one may find a list of references e.g. in a recent paper
\cite{betapert}, which came to the same conclusion via
analysis of higher-loop corrections to $\sigma(e^+e^- \rightarrow
hadrons)$ ). If so, at large $\rho$ the action is $\rho$-independent,
 and the
 following simple prediction follows:
$d(rho)\sim \rho^{-d+1}$, where d is just the space-time dimension.
 Are available data in agreement with it?

   In  terms of statistical accuracy and the widest  range
  studied, the best lattice measurements are those
  for  d=2 O(3) sigma model
\cite{sigma}.
For $large$ sizes the result
is  $d(\rho)\sim\rho^{-3}$ \cite{MS}, in perfect agreement with
the idea of  ``frozen" $S(\rho)$.

  Furthermore, for
  the  d=4 SU(2) gauge
theory (see Fig.1(a)) we have fitted
the action using the following
simple parametrization, a
standard two-loop expression for the charge
(where  $b_0={11\over 3}N_c, b_1={17\over 3}N_c^2$)
\be
{8\pi^2 \over g^2(\rho)}= b_0 L+ b_1 log L
\ee
with a ``regularized" log
\be
L= {1\over p}log[({1 \over \rho\Lambda})^p+ C^p]
\ee
It has two new parameters, C and  p, describing  where and how rapidly the
 ``freezing" occurs. Solid line in Fig.1a shows such a
fit, with
$\Lambda_{inst}$ =.66 $fm^{-1}$, p=3.5, C=4.8.
 For the SU(3) gauge theory the data for $d(\rho)$ itself are still missing,
but  the total instanton density and $average$ size
were measured
in \cite{CGHN_94}:
 $\int d\rho d(\rho)\approx 1.3 fm^{-4},\bar\rho=0.35 fm $.
 One may fix the parameters  to reproduce
those two integrals:
 the corresponding solid curve is shown in Fig.1(b).
(Parameters in this case are $\Lambda_{inst} =.70 fm^{-1}$, p=3.5,
 C=5.0.  ). Although it is $not$ a fit to open points (the interacting liquid)
and is based on different physics, the curve agrees with the points perfectly.
Now we plot both fits it in terms of
the action (Fig.2a) one can
see that it indeed means rather rapid ``freezing" of the coupling constants.
   We do not know why such freezing may
occur: but it is remarkable
that it should happen where the action itself is large, $S\sim
10-20$ (so that one should not question
 the semiclassical theory).
\begin{figure}
\begin{center}
\leavevmode
\epsffile{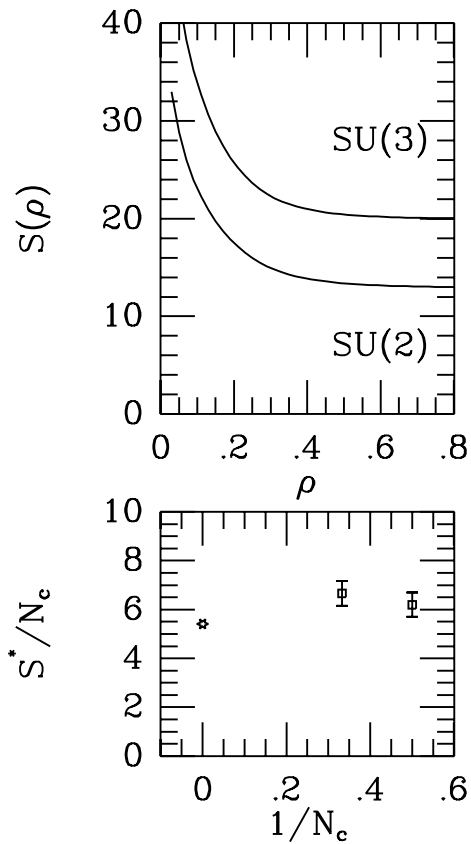}
\end{center}
\caption{The upper part shows
 the instanton action $S(\rho)$ versus $\rho$ (in ``fm")
according to the parametrization used in Fig.1.
The lower part compare the fitted values of the ``fixed point" actions $S^*$
(divided by the number of colors $N_c=2,3$) with  the critical point
in the large $N_c$
limit (star).   }
\end{figure}

   Let us now comment on the alternative (iii) mentioned above,
namely that the cutoff of the large size instantons is related with
$confinement$.
Two arguments $against$
its dominance can be given.
 Although no specific formulae are known
(they should depend on the unknown confinement mechanism), the
confinement-related  correction to the action should  have the form
$\delta S_{conf}(K\rho^2)$ where K is the string tension. Most probably
it is a regular function, expandable in powers of its argument: this leads
 to an $exponential$ cutoff at large
$\rho$ rather than power-like one, as suggested by Fig.1. The second: studies
of the instanton sizes at non-zero temperatures \cite{CS}
have recently  been made. Near the
deconfinement transition (where $K(T)\rightarrow 0$) the
average instanton size does $not$ grow, while one should expect
it to happen  (as $1/K^{1/2}$) if confinement is involved.
(One may investigate
this question further by a number of methods: in particular,
by checking whether
the so called ``abelian projected monopole loops"\cite{Suzuki}
 are or are not correlated
with instantons in the vacuum ensemble.)

   Now, is    the existence of
the infrared fixed point consistent
with non-perturbative beta functions
used in the lattice studies? Recall that the bare coupling g(a)
is fixed at
 the input of simulations, then one
calculates  observables (hadronic masses, etc)
and fixes the $physical$ magnitude of
the lattice scale  a.
 Reversing the function, one gets the renormalized bare charge
  g(a), subject for
(i) universality test (its independence on the  particular  observable
used) and (ii) comparison with  the
 expected perturbative
behaviour (``asymptotic scaling"). The non-trivial fact found
(see  e.g. \cite{Gupta,Blum_etal}) is that (i) extends beyond
(ii).

  A sample of lattice
 data for the  SU(3) lattice gauge theory, without quarks
\cite{Gupta}
(open points) and with two massless quark flavors (closed points)
\cite{Blum_etal} shown in Fig.3. Those are usually presented in form of the
$derivative$ of g(a), the famous Gell-Mann-Low
beta function. In order to show various theories, one can
 normalize the derivative
to its
asymptotic ($g\rightarrow 0$) perturbative value: thus we  plot
 the following ratio
\be
R_{\beta}(g)=[{d(1/g^2) \over dlog(a)}]/[{d(1/g^2) \over
dlog(a)}|_{g\rightarrow 0}
\ee
which tends to 1 the right hand side of Fig.3.
As one penetrates into the non-perturbative region
the data display a deep $drop$ of $R_{\beta}$
  at
$6/g \approx 6$, with subsequent  turn upward. Both are
  very rapid: the latter turn is known in literature as   a transition
 to  a ``strong
coupling" regime. This drop may be an indication
for the {\it infrared
fixed point},  while the upturn is definitely a
lattice artifact. (Rapid turnovers are
 common when a renormalization trajectory is going toward
 the fixed point, and then misses it closely.)
 We conjecture that for lattice actions which are  $better$ than Wilson
one,
the corresponding beta function would drop further.

     However, even with
     $standard$ Wilson actions one may investigate
other definitions of the  renormalized charge, namely
those related with the
renormalized {\it background fields}. This method is quite popular
 in the perturbative context\cite{super}
 but (to our knowledge) has not yet been used
on the lattice.
It is clear   which field is the best to try.
 First, to avoid complications with external current,
one should better take a ``self-supporting" classical field, such as
  $D_\mu G^{c}_{\mu\nu}=0$. Second,
topology adds additional stability: thus $instanton$ is the most
 natural
classical background. Third,
its renormalized action depend on one parameter $\rho$,
$S_{eff}=8\pi^2/g^2(\rho)$, and thus this expression
can be used to study the non-perturbative
charge renormalization.
\begin{figure}
\begin{center}
\leavevmode
\epsffile{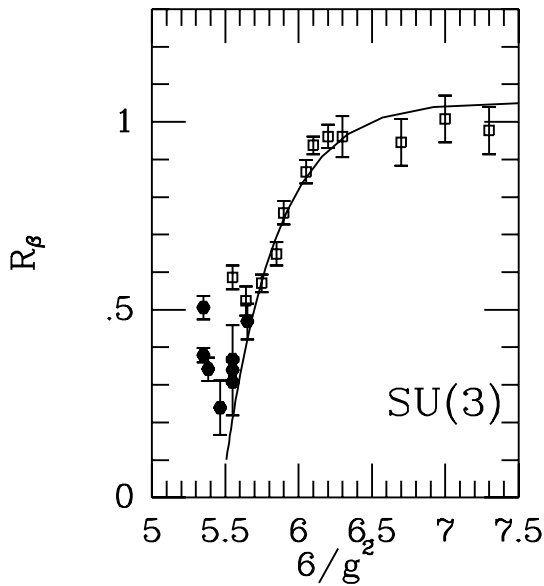}
\end{center}
\caption{
The non-pertubative beta function for the SU(3) lattice theory with Wilson
action. Open points are taken from R.Gupta review  (which is  an
analysis of the data set from Lapage and Mackenzie), the solid points
are from Blum et al. The solid line is a fit discussed in the text.
}
\end{figure}

   One can put an instanton on the lattice and then ``heat it up",
 performing standard
updates:  such studies have been made by DiGiacomo et al \cite{DiGiacomo},
establishing
renormalization of the topological susceptibility. However, for the proposed
 goal is not enough
 to keep a topological charge  (as done in
ref.\cite{DiGiacomo}):  one has to
preserve the chosen value of $\rho$. This
can be
achieved in two ways: (i) while updating a link, one may  keep the
quantum field $a_\mu$  $orthogonal$
to the $dilatational$ zero mode $\delta A^c_\mu/\delta \rho$
\cite{AS}; or (ii) one may
use a modified lattice action containing the  two-plaquette operators with
parameters tuned to make any given
size  the classical minimum of the lattice action.
Either way, the main problem is to
get high statistics measurements of the effective action, after subtraction
of the usual ``average plaquette" is made.

   The next topic addressed in this Letter is the fate of instantons in the
large $N_c$ limit. It was first addressed by Witten
\cite{Witten}, who has formulated the  following
 dilemma: either (i)
 instantons are not dynamically
relevant, while all observables have
 regular $1/N_c$ expansion
as suggested by perturbative expansion, or  (ii)  instantons
 play some role in the real world, but are
exponentially suppressed at large $N_c$.
Back in 1979, Witten
has argued in favor of (i), using analogies
 with some d=2 models, but today there is
no doubt about
 significance of the instanton-induced effects at $N_c=3$. However,
it is hard to accept (ii) also, because  then all
the $1/N_c$ development is undermined.

   This dilemma still allows for one loophole
\cite{Shuryak_1982}, an alternative (iii),
 which allows to reconcile
 instanton physics with smooth large $N_c$ limit.
Let us look at its consequences
and compare them  with the empirical trends found for $N_c=2,3$.
 The semiclassical formula (2)
in the large $N_c$ limit contains $factorial$ terms in
the denominator. The action
 should grow $linearly$ with $N_c$:  $S(\rho)=N_c s(\rho)$
and therefore these
factorials  are exactly cancelled by  the gauge zero modes. The next level are
the $exponential$ terms, which with our normalization look as follows
\be
d(\rho) \sim exp \{N_c[2-s(\rho)+2 log s(\rho)] \}
\ee
The limit clearly depends on the sign of the bracket.
If  there exist the fixed point $s(\rho)\rightarrow s^*$,
as we advocated above, it is important whether it is larger
or smaller than the {\it critical value} $s_c^*=5.4$, the root
 of the bracket.

  Let us
 compare the values of the fixed actions for $N_c=2,3$ obtained from the
fits above to the critical
 value (shown by the star) in Fig.2(b).
The first observation is that
for two fitted cases the limiting action $S^*$ indeed grow
$linearly$ with $N_c$. Second, the constant is somewhat
 $larger$ than the critical
value, although this is so to say inside
the error bars (we have assign those to the points,
which relate to the uncertainty due to instanton interactions).
In general, it seems quite plausible that at $N_c\rightarrow \infty$ instantons
will take care of themselves and their density and role will remain roughly
constant.

  But should we expect the limit $N_c\rightarrow \infty$ to be smooth?
The answer is $negative$    \cite{DP}:
  at  larger $N_c$ an ``instanton
liquid" should become  solid, because a
 growing action (which depends on relative positions) is analogous to
a decreasing temperature. Thus, one should expect that
 both color and translational
symmetries would be spontaneously broken.
 In \cite{SV} such phase transition
was in fact found for $N_c=3$, but at $non-physical$
 instanton density (about 60 times
the physical one). For larger $N_c$ the work is in progress.
Lattice studies of theories with $N_c>3$
would certainly be of great value.

   Finally, let us mention the practical aspect of the studies of $d(\rho)$:
 its measurements
 at $small$ $\rho<0.2 fm$ is potentially the source of
by far the most accurate measurements of $\Lambda_{QCD}$. As it is well known,
here $d(\rho) \sim
\Lambda^{b_0}_{QCD}\rho^{(b_0-d-1)}$: thus
even relatively poor accuracy
(say 50\%, compare to dots at Fig.1a) leads to
$\Lambda_{QCD}$  with accuracy of about 5\%, much better
than measured today.

  In summary,  we have argued that
existing data for the instanton size distribution can be equally well be
described both by the interacting instanton liquid model (IILM) and by
the existence of
the infrared fixed point. If the second is the case, it seems plausible that
this fixed point is not far from the critical point, separating
scenarios in which
instantons are suppressed or grow in the $N_c\rightarrow \infty$ limit.
Better measurements of $d(\rho)$ on the lattice (especially at $N_c=3$ and
$larger$) are badly needed.

{\bf Acknowledgements} This work is partially supported by the DOE grant
DE-FG02-88ER40388.


\begin{thebibliography}{100}
\bibitem{sigma}Recall that  this model is also asymptotically free
with the one-loop beta
function coefficient $b_0$=2.
 Thus the semiclassical theory leads to
 $d(\rho)\sim 1/\rho$ at small $\rho$, or
ultraviolet divergent
total density and topological susceptibility.
\end{thebibliography}
\end{document}